\documentclass[12pt]{article}
\usepackage[pctex32]{graphics}
\begin{document}

\vspace{2cm}
\begin{center} {\large \bf  Condensation of  Tubular D2-branes  in  Magnetic Field Background}
                                                  
\vspace{1cm}

                      Wung-Hong Huang\\
                       Department of Physics\\
                       National Cheng Kung University\\
                       Tainan,70101,Taiwan\\

\end{center}
\vspace{1cm}
\begin{center} {\large \bf  Abstract} \end{center}
It is known that in the Minkowski vacuum a bunch of IIA superstrings with D0-branes can be blown-up to a supersymmetric tubular D2-brane, which is supported against collapse by the angular momentum generated by crossed electric and magnetic  Born-Infeld (BI) fields.   In this paper we show how the multiple, smaller tubes with relative angular momentum could condense to a single, larger tube to stabilize the system.   Such a phenomena could also be shown in the systems under the Melvin magnetic tube or  uniform magnetic field background.    However, depending on the magnitude of field strength, a tube in the uniform magnetic field background may split into multiple, smaller tubes with relative angular momentum to stabilize the system.

\vspace{4cm}
\begin{flushleft}
E-mail:  whhwung@mail.ncku.edu.tw\\
\end{flushleft}
\newpage
\section {Introduction}
In an interesting paper [1], Mateos and Townsend showed that a cylindrical, or tubular D2-brane in a  Minkowski vacuum spacetime can be supported against collapse by the angular momentum generated by crossed electric and magnetic  Born-Infeld (BI) fields.   The background in this case is trivial as there is no external force as that in the Myers effect [2], in which the D0 branes in an external RR four-form field will expand into the fuzzy sphere to stabilize the system. 

    In this paper we will first show how the multiple, smaller tubes with relative angular momentum could condense to a single, larger tube to stabilize the system in a  Minkowski vacuum spacetime.   We next investigate the phenomena in the system under the Melvin magnetic tube  background.  Note that the Melvin metric is a solution of  Einstein-Maxwell theory, which describes a static spacetime with a cylindrically symmetric magnetic flux tube.   It provides us with a curved space-time background in which the superstring theory can be solved exactly [4].   In the Kaluza-Klein spacetime the Melvin solution is a useful metric to investigate the decay of magnetic field [5] and the decay of spacetimes, which is related to the closed string tachyon condensation [6].   The fluxbranes in the Melvin spacetime have many interesting physical properties as investigated in the resent literatures [7].   

   We also investigate the phenomena in the system under a uniform magnetic field background.  The background can be  obtained by Kaluza-Klein reduction of a special class of plane wave solutions and also provides us with a curved space-time background in which the superstring theory can be solved exactly [8,9].   The problem of the decay of the uniform magnetic field, which is related to the closed string tachyon condensation had also been studied in [10].  We also see that, in the uniform magnetic field background, the multiple, smaller tubes could condense to a single, larger tube to stabilize the system. However, depending on the magnitude of field strength, a tube in the uniform magnetic field background may split into multiple, smaller tubes with relative angular momentum to stabilize the system.

\section {Tube in Melvin Magnetic Tube Field Background}

  In the ten-dimensional IIA background the Melvin metric is described by [3]
$$ ds_{10}^{2} = \Lambda^{1/2}\left(-dT^2+\sum_{m=1}^{6}dy_mdy^m + dZ^2 + dR^2\right)  +\Lambda^{-1/2}R^2d{\Phi}^2 ,  $$
$$e^{4\phi/3}=\Lambda  ,~~~~ A_{\Phi}=\frac{Br^2}{2\Lambda}, ~~~~~~~\Lambda \equiv 1+ R^2 f^2 ,   \eqno{(2.1)} $$
The parameter $f$ is the magnetic field along the $Z$-axis defined by
$f^2=\frac{1}{2}F_{\mu\nu}F^{\mu\nu}|_{\rho=0}$.  The Melvin spacetime is  an exact solution of M-theory and can be used to describe the string propagating in the magnetic tube field background.

    The D2-brane Lagrangian, for unit surface tension, is written as 
$${\cal L} =  - e^{-\phi}~\sqrt{- \det (g +F)},   \eqno{(2.2)} $$
where $g$ is the induced worldvolume 3-metric and $F$ is the BI 2-form
field strength.     If we take the worldvolume coordinates to be $(t,z,{\varphi})$ with ${\varphi} \sim {\varphi} + 2\pi$, then we may fix the
worldvolume diffeomorphisms for a D2-brane of cylindrical topology by
the `physical' gauge choice
$$T=t ,  ~~~~ Z = z , ~~~~~~ \Phi= \varphi .  \eqno{(2.3)} $$
For a static cylindrical D2-brane of radius $R$, with the $z$-axis as the axis of symmetry, the induced metric is
$$ ds^2(g) =\Lambda^{1/2}\left(-dt^2+ dz^2\right)  +\Lambda^{-1/2}R^2d{\varphi}^2 .\eqno{(2.4)} $$
We will allow for a time-independent electric field $E$ in the $z$-direction, and a time-independent magnetic field $B$,  so the BI 2-form field strength is [1]
$$ F= E \, dt\wedge dz + B \, dz\wedge d\varphi . \eqno{(2.5)}$$
The field will generate an angular momentum to stabilize the tubular D2 brane and prevent its collapsing to zero radius [1]. Under these conditions the Lagrangian becomes
$$ {\cal L} = - \Lambda^{-3/4}\sqrt{\Lambda^{1/2} (R^2 +B^2)- E^2 R^2 \Lambda^{-1/2}}.  \eqno{(2.6)}$$
The momentum conjugate to $E$ takes the form
$$\Pi \equiv {\partial{\cal L}\over \partial E}  =  {E R^2 \Lambda^{-5/4}\over \sqrt{\Lambda^{1/2} (R^2 +B^2)- E^2 R^2 \Lambda^{-1/2}}}.   \eqno{(2.7)}$$ 
The corresponding Hamiltonian density is 
$$ {\cal H} \equiv \Pi E - {\cal L} = {\Lambda^{-1/2}\over R} \sqrt{(R^2+B^2)(R^2 + \Lambda^{1/2} \Pi^2)}~.   \eqno{(2.8)}$$
 For an appropriate choice of units, the integrals 
$$q_s \equiv {1\over 2\pi}\oint d\varphi \, \Pi \qquad \mbox{and} \qquad 
q_0 \equiv {1\over 2\pi} \oint d\varphi \, B   \eqno{(2.9)}$$
are, respectively, the conserved IIA string  charge and the D0-brane charge per unit length carried by the tube [1].  

We can now use the Hamiltonian density (2.8) to perform the following two analyses.  Let us first see the system in the Minkowski vacuum spacetime.   In this case (2.8) become 
$$ {\cal H}(R, B, \Pi) = {1\over R} \sqrt{(R^2+B^2)(R^2 + \Pi^2)}.   \eqno{(2.10)}$$
The Hamiltonian density has a minimum at finite radius $R_0 = \sqrt{B\Pi}$ and 
$$ {\cal H}_{min}(R, B, \Pi) = |B| + |\Pi| = |q_s| + |q_0|.  \eqno{(2.11)}$$
Therefore we have a simple relation 
$${\cal H}_{min}(R, NB, N\Pi) = N~{\cal H}_{min}(R, B, \Pi).  \eqno{(2.12)}$$
This relation tells us that a big tube, formed from the IIA superstring with charges $Nq_s$ and carrying D0-brane charges $Nq_0$ has the same energy as N smaller tubes, formed from the IIA superstring with charges $q_s$ and carrying D0-brane charges $q_0$.   However, the big tube will carry angular momentum $J_{big}= N^2q_s q_0$ which is larger than the angular momentum $J_{N_ {small}}= N q_s q_0$ carrying by the $N$ smaller tubes.    Thus, the $N$ smaller tubes moving with relative angular momentum $J_{rel}= (N^2 -N) q_s q_0$, which will add to the over all energy of the tubes [11] and break the supersymmetry, will condense to a single, larger tube to stabilize the system [12]. 

For example, in figure 1 we show the dynamical process of joining two tubes. First, two tubes with a relative angular momentum approach to each other.  After hitting on the two tubes will join and recombine [13] itself to a new  configuration.  Finally, a larger tube of  stable configuration is formed.
\\
\\
\scalebox{0.7}{\includegraphics{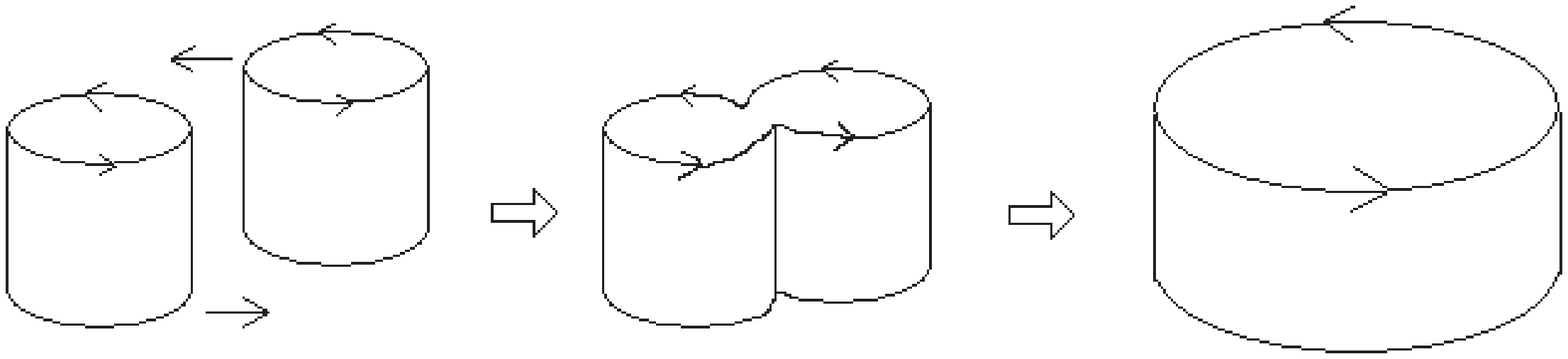}}\\
\hspace{2cm}{\it ~~~Figure.1. The dynamical process of joining two tubes.   Two tubular configurations with relative angular momentum approach to each other, then they hit on, recombine, and join to a larger tube.}\\
\hspace{1cm}

There is an another possible situation in which two tubes hit on to each other while without relative angular momentum.   In this case the recombined configuration might be a joined two-cylinders of the shape of the Arabic letter $8$  \footnote {This dynamical joining and splitting of the tubes is suggested by the referee.}.  However, as the energy of tube will depend on the cross sectional shape [11] it seems difficult to analyze the problems therein.  Thus, in this paper we will consider only the case of the tubes with circular cross section, which have a relative angular momentum.

The important constraints in the process of condensation are the conservations of the charge and angular momentum.  The above analyses tell us that to satisfy the two constraints the $N$ smaller tubes shall be moving with suitable relative angular momentum to condense to a single, larger tube.    

   Another possible phenomena is that the multiple, static tubes could also condense to a single tube.    In this case, however, to satisfy the two constraints the angular momentum of the constituent static sub-tubes shall not all at the same direct.   To see this fact let us consider two tubes with charges $Q_1 \equiv (q_0, q_s) = (B, \Pi_1)$, $Q_2 = (- B ~{\Pi_2\over \Pi_1}, \Pi_2)$ and angular momenta $J_1= B \Pi_1$,  $ J_2= -B  \Pi_2 ~{\Pi_2\over \Pi_1}$, respectively ($B$ and $\Pi_i$ are chosen to be positive.).     The two tubes have total charge $Q_{1+2}=( B - B ~{\Pi_2\over \Pi_1}, \Pi_1 +\Pi_2)$. and total energy $ E_{1+2}= \Pi_1 + \Pi_2 + B + B ~{\Pi_2\over \Pi_1}$.   A single tube we consider has charges $Q = Q_{1+2}$, which will have angular momentum $J= (\Pi_1 + \Pi_2) (B - B ~{\Pi_2\over \Pi_1})$ and energy $E = \Pi_1 + \Pi_2 + |B - B ~{\Pi_2\over \Pi_1}|$.    It can be seen that the constraints $Q = Q_1+Q_2$ and $J= J_1+J_2$ are satisfied.   However, as $E_1 + E_2 > E$ the two tubes will condense to a single tube. 

   We now turn to the system in the external magnetic field background.  The Hamiltonian density (2.8) is 
$$ {\cal H}  = {\left(1+f^2 R^2\right)^{-1/4}\over R} \sqrt{(R^2+B^2)\left(\left(1+f^2 R^2\right)^{-1/2}R^2 + \Pi^2\right)},   \eqno{(2.13)}$$
which has a minimum at a finite radius.    The Hamiltonian has two interesting properties
$${\cal H}(R, f, NB, N\Pi) = N~{\cal H}\left({R\over N}, Nf,  B, \Pi\right).  \eqno{(2.14)}$$
$${\cal H}\left(R, Nf,  B, \Pi \right)  < {\cal H}\left(R, f,  B, \Pi \right), ~~~~~~  if ~~N > 1 .\eqno{(2.15)}$$
Using the properties we have the relation
$${\cal H}_{min}(R, f, NB, N\Pi) = N~{\cal H}_{min}\left({R\over N}, Nf,  B, \Pi\right)  < N~{\cal H}_{min}\left({R\over N}, f,  B, \Pi\right),~~~~~~if ~~N > 1. \eqno{(2.16)}$$
Because that 
$${\cal H}_{min}\left({R\over N}, f,  B, \Pi\right) = {\cal H}_{min}\left(R, f,  B, \Pi\right), \eqno{(2.17)}$$
we have the final relation 
$${\cal H}_{min}(R, f, NB, N\Pi) < N~{\cal H}_{min}\left(R, f,  B, \Pi\right). ~~~~~~if ~~N > 1, ~~~~ ~f \ne 0     \eqno{(2.18)}$$
This equation implies that in the Melvin magnetic tube background the N tubes with charge $Q = (B, \Pi)$  will have higher energy then a bigger tube which has charge $Q = (NB, N\Pi)$.   Then, as the discussions in the Minkowski vacuum the, $N$ smaller tubes which are moving with suitable relative angular momentum will to condense to a single, larger tube. 

\section {Tube in Uniform Magnetic Field Background}
The metric of ten-dimensional spacetime with uniform magnetic field is described by [8-10]
$$ ds^2  =  -\left (dt + \sum_{i=1}^{2} \epsilon_{ij}  \frac{f }{2} x^j dx^i\right)^2 + \sum_{i=1}^{2} dx^idx_i + \sum_{m=3}^{9}dx^m dx_m,  \eqno{(3.1)}$$
in which $f$ is the strength of the magnetic field and the dilaton is constant.   This solution is different from  the Melvin background in which the magnetic field strength decreases from a finite value at the origin to zero at infinity.

    As in section 2, we allow for a time-independent electric field $E$ in the $z$-direction, and a time-independent magnetic field $B$.       For a static cylindrical D2-brane of radius $R$ the Lagrangian, for unit surface tension, is 
$${\cal L}= -\sqrt{R^2 (1 - E^2) +(B - f E R^2/2)^2}.  \eqno{(3.2)}$$
The corresponding Hamiltonian density becomes 
$${\cal H}= {1\over R(f^2 R^2-4)}\left(2\Pi B f R - \sqrt{(4B^2 + 4R^2 - f^2 R^4)(4\Pi^2 + 4R^2 - f^2 R^4)}\right),  \eqno{(3.3)}$$
in which $\Pi$ is the momentum conjugate to $E$.   The Hamiltonian has a minimum at a finite radius.    Thus strings carrying D0-brane charges can be blown-up to a  tubular D2-brane with a finite radius.   The energy of the tubular D2 brane is the function of  the string charge $q_s$, D0-brane charge $q_0$, and strength of the background magnetic field $f$. 

   To proceed, we will consider the system with $\Pi = B$.   In this case the Hamiltonian density becomes 
$${\cal H}={2B^2 \over R(fR+2)} + R,  \eqno{(3.4)} $$
which has a simple form and we could therefore perform the following analyses.     
  
 The Hamiltonian density (3.4) has the interesting properties
$${\cal H}(R, f, NB) = N~{\cal H}\left({R\over N}, Nf,  B\right).  \hspace{3.5cm}\eqno{(3.5)}$$
$${\cal H}\left(R, Nf,  B \right) ~ {}^> _{<}~{\cal H}\left(R, f,  B \right), ~~~~~~  if ~~N > 1, ~~~~ f~  {}^< _{>} ~0. \eqno{(3.6)}$$
Using the properties we have the relation
$${\cal H}_{min}(R, f, NB) = N~{\cal H}_{min}\left({R\over N}, Nf,  B\right)  ~ {}^> _{<}~ N~{\cal H}_{min}\left({R\over N}, f,  B\right),~~~  if ~~N > 1, ~~ f~  {}^< _{>} ~0. \eqno{(3.7)}$$
As we also have a property ${\cal H}_{min}\left({R\over N}, f,  B\right) = {\cal H}_{min}\left(R, f,  B\right)$,  we have the final relation 
$${\cal H}_{min}(R, f, NB) ~ {}^> _{<}~ N~{\cal H}_{min}\left(R, f,  B\right),~~~~~~  if ~~N > 1, ~~~~ f~  {}^< _{>} ~0. \eqno{(3.8)}$$
The case of $f>0$ will behave as before as can be easily seen.  So let us turn to the case of $f<0$. 

   As before, to satisfy the conservation of angular momentum we may consider the case of the N moving tubes with relative angular momentum.  The additional energy coming from the relative moving is $E_{rel} = {J_{rel}^2 \over 2 I}$ , in which $I$ is the inertial moment of the N tubes and $J_{rel}$ is the additional angular momentum from the relative moving.   In the case of very large inertial moment  (for example, the distances between constitute tubes are very large) the energy $E_{rel}$ will be very small and thus the N tubes could still have less energy then that of a bigger tube.   In this case, a tube in the uniform magnetic field background may split into multiple, smaller tubes with relative angular momentum to stabilize the system.

   In general, the system may show the condensation of multiple tubes or the splitting of a large tube, depending on the magnitudes of the $\Pi$ ($\sim$ string charges $q_s$), $B$ ($\sim$ D0 charges $q_0$) and $f$ (the strength of magnetic field). 

\section {Conclusion}

   In conclusion,  a cylindrical, or tubular D2-brane in a Minkowski vacuum spacetime can be supported against collapse by the angular momentum generated by crossed electric and magnetic  Born-Infeld (BI) fields.  A bunch of IIA superstrings with D0-branes can thus be blown-up to a supersymmetric tubular D2-brane.  In this paper we show how the multiple, smaller tubes with  relative angular momentum could condense to a single, larger tube to stabilize the system.   We also show that such a phenomena could be seen in the system under the Melvin magnetic tube or  uniform magnetic field background.   Especially, we find that, depending on the magnitude of field strength, a tube in the uniform magnetic field background may split into multiple, smaller tubes with relative angular momentum to stabilize the system.

   Finally, the dynamics of splitting and joining tubes may also be investigated in the matrix model [14,15] or the effective tachyon action [16,17].  We will study the problems in the future papers.

\newpage
\begin{center} {\large \bf  References} \end{center}
\begin{enumerate}
\item D. Mateos and P. K. Townsend, ``Supertubes'', Phys. Rev. Lett. 87 (2001) 011602 [hep-th/0103030];  R. Emparan, D. Mateos and P. K. Townsend,``Supergravity Supertubes'', JHEP 0107 (2001) 011 [hep-th/0106012]; D.~Mateos, S.~Ng and P.~K.~Townsend, ``Tachyons, supertubes and brane/anti-brane systems'', JHEP  0203 (2002) 016 [hep-th/0112054]; Y. Hyakutake and N. Ohta, ``Supertubes and Supercurves from M-Ribbons,'' Phys. Lett. B539  (2002) 153 [hep-th/0204161]; N. E. Grandi and A. R. Lugo , `` Supertubes and Special Holonomy'',  Phys. Lett. B553 (2003) 293 [hep-th/0212159];
\item R.C. Myers, ``Dielectric-Branes'',  JHEP 9912 (1999) 022 [hep-th/99010053].  
\item M.A. Melvin, ``Pure magnetic and electric geons,'' Phys. Lett. 8 (1964) 65; G.~W.~Gibbons and D.~L.~Wiltshire, ``Space-time as a membrane in higher dimensions,'' Nucl.\ Phys.\ B287 (1987) 717 [hep-th/0109093]
\item  J.~G.~Russo and A.~A.~Tseytlin, ``Exactly solvable string models of curved space-time backgrounds,'' Nucl.\ Phys.\ B449 (1995) 91 [hep-th/9502038];``Magnetic flux tube models in superstring theory,'' Nucl.\ Phys.\ B461 (1996) 131 [hep-th/9508068].
\item F.~Dowker, J.~P.~Gauntlett, D.~A.~Kastor and J.~Traschen, ``The decay of magnetic fields in Kaluza-Klein theory,'' Phys.\ Rev.\ D52 (1995) 6929 [hep-th/9507143];  M.~S.~Costa and M.~Gutperle, ``The Kaluza-Klein Melvin solution in M-theory,'' JHEP 0103 (2001) 027 [hep-th/0012072].
\item  A. Adams, J. Polchinski and E. Silverstein,  ``Don't Panic! Closed String Tachyons in ALE Spacetimes,''  JHEP 0110 (2001) 029 [hep-th/0108075]; J.~R.~David, M.~Gutperle, M.~Headrick and S.~Minwalla, ``Closed string tachyon condensation on twisted circles,''  JHEP 0202 (2002) 041 [hep-th/0111212]; T.~Takayanagi and T.~Uesugi, ``Orbifolds as Melvin geometry,''   JHEP 0112 (2001) 004 [hep-th/0110099].
\item M.~Gutperle and A.~Strominger, ``Fluxbranes in string theory,''
JHEP 0106 (2001) 035 [hep-th/0104136];  R.~Emparan and M.~Gutperle, ``From p-branes to fluxbranes and back,'' JHEP 0112 (2001) 023 [hep-th/0111177]; T.~Takayanagi and T.~Uesugi,  ``D-branes in Melvin background,'' JHEP 0111 (2001) 036 [hep-th/0110200].
\item  J.~G. Russo and A.~A. Tseytlin, ``Constant magnetic field in closed string theory: An exactly solvable model,'' Nucl. Phys.  B448 (1995)  293 [hep-th/9411099].
\item  J.~G. Russo and A.~A. Tseytlin, ``Heterotic strings in uniform magnetic field,'' Nucl. Phys. B454 (1995) 164 [hep-th/9506071].
\item J. David, ``Unstable magnetic fluxes in heterotic string theory,''  JHEP 0209 (2002) 006 [hep-th/0208011].
\item D. Bak, Y. Hyakutake, and N. Ohta, ``Phase Moduli Space of Supertubes,''  [hep-th/0404104].
\item After releasing the manuscript on arXiv  I receive an E-Mail  from Dr. Eric Gimon, in which a critical flaw of neglecting the angular momentum in the old version is mentioned to me.  The revised version presents a possible solution by letting the constituent sub-tubes move with a relative angular momentum, as suggested by Eric Gimon. 
\item  K. Hashimoto and S.~Nagaoka, ``Recombination of intersecting D-branes by local tachyon  condensation,'' JHEP  0306 (2003) 034 [hep-th/0303204];\\
W. H. Huang, ``Recombination of intersecting D-branes in tachyon field theory,'' Phys. Lett. B  564 (2003), 155 (2003) [hep-th/0304171];  W. H. Huang, ``On Tachyon Condensation of Intersecting Noncommutative Branes in M(atrix) theory,''  Phys.Lett. B578 (2004) 418-424 [math-ph/0310005];\\
T. Sato , ``D-brane Dynamics and Creations of Open and Closed Strings after Recommbination,'' Nucl.Phys. B682 (2004) 117 [hep-th/0304237]\\
J. Erdmenger, Z. Guralnik, R.  Helling, I. Kirsch, ``A World-Volume Perspective on the Recombination of Intersecting Branes,''  JHEP 0404 (2004) 064 [hep-th/0309043]\\
S. Nagaoka, `` Higher Dimensional Recombination of Intersecting D-branes,''  JHEP 0402 (2004) 063 [hep-th/0312010]
\item D. Bak, K. M. Lee, ``Noncommutative Supersymmetric Tubes'',  Phys. Lett. B509 (2001) 168 [hep-th/0103148]; D. Bak and S. W. Kim, ``Junction of Supersymmetric Tubes,'' Nucl. Phys.  B622 (2002) 95 [hep-th/0108207]
\item D. Bak and A. Karch, ``Supersymmetric Brane-Antibrane Configurations,'' Nucl. Phys.  B626 (2002) 165 [hep-th/011039]; D. Bak and N. Ohta, ``Supersymmetric D2-anti-D2 String,'' Phys. Lett.  B527 (2002) 131 [hep-th/0112034]; D. Bak, N. Ohta and M. M. Sheikh-Jabbari, ``Supersymmetric Brane-Antibrane Systems: Matrix Model Description, Stability and Decoupling Limits,'' JHEP  0209 (2002) 048 [hep-th/0205265].
\item A. Sen, ``Dirac-Born-Infeld Action on the Tachyon Kink and Vortex,''  Phys. Rev. D68 (2003) 066008 [hep-th/0303057].
\item C. Kim, Y. Kim, O-K. Kwon, and P. Yi, ``Tachyon Tube and Supertube,''  JHEP 0309 (2003) 042 [hep-th/0307184];  Wung-Hong Huang, ``Tachyon Tube on non BPS D-branes,'' [hep-th/0407081]

\end{enumerate}
\end{document}